\documentclass[twocolumn, pra, showpacs, showkeys]{revtex4}
\usepackage{graphicx}
\usepackage{dcolumn}
\usepackage{bm}
\usepackage{amsmath} 

\renewcommand {\vec} {\mathbf}

\begin{document}

\preprint{APS/123-QED}

\title{Coherent spin mixing dynamics in a spin-1 atomic condensate}
\author{Wenxian Zhang$^1$, D. L. Zhou$^{1,2}$, M. -S. Chang$^1$, M. S.
Chapman$^1$, and L. You$^{1,2}$}
\address{$^1$School of Physics, Georgia Institute of Technology,
Atlanta, GA 30332-0430}
\address{$^2$Institute of Theoretical Physics, The Chinese
Academy of Sciences, Beijing 100080, China}
\date{ Submitted on August 13, 2004}

\begin{abstract}
We study the coherent off-equilibrium spin mixing inside
an atomic condensate. Using mean field theory and
adopting the single spatial mode approximation (SMA), the condensate
spin dynamics is found to be well described by that of a nonrigid pendulum,
and displays a variety of periodic oscillations in an
external magnetic field. Our results illuminate
several recent experimental observations and provide critical insights
into the observation of coherent interaction-driven oscillations
in a spin-1 condensate.
\end{abstract}



\maketitle
\narrowtext


Bose-Einstein condensation (BEC) has been one of the most active
topics in physics for over a decade, and yet interest in this
field remains impressively high. Recent experiments showcase the
rich versatility of control over the atomic superfluid, e.g. the
BEC-BCS crossover \cite{Regal04, Zwierlein04}, quantized vortices
\cite{Matthews99vortex, Madison00, Raman01}, condensates in
optical lattices \cite{opticallattice}, and low dimensional
quantum gases \cite{Olshanii98, Gorlitz01}. While most of these
efforts involve condensates of atoms in a single Zeeman state,
activities in spinor condensates \cite{Stenger98, Stamper-Kurn98}
have recently received significant boost with the addition of
three new spin-1 experiments \cite{Barrett01, Schmaljohann04,
Schmaljohann04a, Kuwamoto04}.

In a spinor condensate, atomic hyperfine spin degree of freedom becomes
accessible with the use of a far-off resonant optical trap instead of a
magnetic trap. For atoms in the $F=1$ ground state manifold, the presence of
Zeeman degeneracy and spin dependent atom-atom interactions \cite{Ho98, Ohmi98,
Law98, Pu99, Yi02, Stamper-Kurn98, Barrett01, Stenger98} leads to interesting
condensate spin dynamics. In this article, we study spin mixing inside a spin-1
condensate \cite{Law98, Pu99, Pu00}, focusing on the interaction-driven
coherent oscillations within a mean field description. Unlike the pioneering
studies on this subject as in Refs. \cite{Law98,Pu99}, we will highlight the
important role of an external magnetic field, which is present in all
experiments to date.

Recently, a beautiful experiment has finally observed the long predicted
Josephson type coherent nonlinear oscillations with a scalar condensate in a
spatial double well potential \cite{Albiez04}.

Although spin mixing driven by the internal spin-dependent interaction (not of
the nature of a Rabi oscillation as driven by an external field
\cite{Matthews99, Williams00}), has been observed in both $F=1$ and $F=2$
condensates \cite{Stenger98, Schmaljohann04, Chang04, Kuwamoto04}, the
coherence of this process has not yet been investigated. Over-damped single
oscillations in spin populations have been observed in earlier experiments
\cite{Chang04} although their interpretation has been limited because evolution
from the initial (meta-stable) states was noise-driven. The main experimental
obstacles to observe more oscillations are the dissipative atomic collisions
among the condensed atoms and the decoherence collisions with noncondensed
atoms \cite{Schmaljohann04, Chang04}. A promising future direction relies on
increased atomic detection sensitivity, thus the use of smaller condensates as
in the experiment of Ref. \cite{Albiez04}, with lower number densities and at
lower temperatures, two favorable conditions for the single spatial mode
approximation (SMA) \cite{Law98, Pu99}.

\begin{figure}
\includegraphics[width=3.25in]{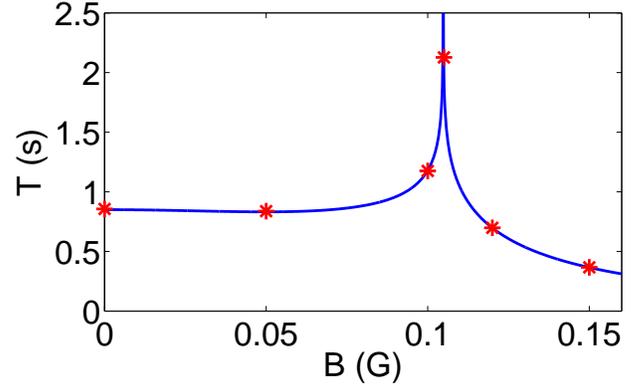}
\caption{(Color online) The dependence of oscillation period $T$ on magnetic
field $B$ for a $^{87}$Rb condensate from our model (solid line, Eq.
(\ref{RbTsma})), the results from a full numerical simulation without the use
of SMA are denoted by (*). } \label{fig1}
\end{figure}

The initial atomic population distribution in Fig. \ref{fig1} corresponds to
the (equilibrium) ground state at a magnetic field (B-field) of $0.07$ Gauss
and with a zero magnetization ($m=0$), specified by $\rho_0(0)\approx 0.644$
and $\theta(0)=0$ with $c\approx 0.614$ Hz (these symbols are defined
later). As in the case of no B-fields \cite{Law98, Pu99, Yi02}, the initial
relative phases among the three components depend on the spin-dependent
atom-atom interaction being ferromagnetic ($0$) or antiferromagnetic ($\pi$),
inside an external magnetic field \cite{NJP}. Starting with the initial phases
and population distributions at $B_0=0.07$ Gauss, we instantaneously change the
$B$-field to a different value, the atomic condensate distributions thus become
off-equilibrium, and the coherent dynamics starts according to the mean field
theory. Within the SMA, we find such off-equilibrium dynamics of a spin-1
condensate corresponds to that of a nonrigid pendulum, which can be
characterized using semiclassical trajectories in the phase space.

Our main result is illustrated in Fig. \ref{fig1}, where we have plotted the
dependence of oscillation period on the external magnetic field. The parameters
are close to the experiment \cite{Chang04a}; where the spin-independent trap is
harmonic $V = (M/2)(\omega_x^2x^2 +\omega_y^2y^2+\omega_z^2z^2)$ with
$\omega_x=\omega_y= (2\pi) 240$ Hz and $\omega_z=(2\pi) 24$ Hz. The condensate
contains $N=1,000$ $^{87}$Rb atoms with an average density of $\langle n\rangle
\approx 1.7\times 10^{13}$ cm$^{-3}$.

As illustrated in Fig. \ref{fig1}, the spin mixing dynamics within the SMA
corresponds to a typical pendulum, with the quadratic Zeeman energy playing the
important role of the total energy. At small change of $B$-field, the
equivalent pendulum undergoes a small amplitude oscillation, approximately
harmonic with a period independent of the energy or oscillation amplitude;
increasing of the total energy leads to a longer oscillation period as the
pendulum becomes increasingly nonlinear. At a critical field $B_c$, when the
effective total energy is just enough to bring the pendulum to the completely
up or top position, the period approaches infinity as for the homoclinic orbit
of a pendulum; Upon further increasing the energy (or $B$), the pendulum starts
to rotate around and the period becomes smaller with increasing energy as the
pendulum rotates faster and faster.

Our system of a spin-1 atomic Bose gas inside an external magnetic field
is described by the Hamiltonian \cite{Ho98, Ohmi98}
\begin{eqnarray}
{\cal H} &=& \int d\vec r\, \left[\psi_i^\dag
    \left(-{\hbar^2\over 2M}\nabla^2 + V + E_i\right) \psi_i
    +{c_0\over 2} \psi_i^\dag\psi_j^\dag\psi_j\psi_i \right.
    \nonumber\\
 & &\left.+{c_2\over 2} \psi_k^\dag\psi_i^\dag\left(F_\gamma
    \right)_{ij}\left(F_\gamma \right)_{kl}\psi_j \psi_l \right],
\label{h}
\end{eqnarray}
where repeated indices are summed, and
$\psi_i(\vec r)$ ($\psi_i^\dag$) is the field operator that
annihilates (creates) an atom in the $i$-th hyperfine
state ($|F=1,i=+1,0,-1\rangle$, hereafter $|i\rangle$)
at location $\vec r$. $M$ is the mass of an atom.
Interaction terms with
coefficients $c_0$ and $c_2$ describe respectively elastic collisions
of spin-1 atoms, expressed in terms of the scattering
length $a_0$ ($a_2$) for two spin-1 atoms in the combined symmetric
channel of
total spin $0$ ($2$), $c_0=4\pi\hbar^2(a_0+2a_2)/3M$ and
$c_2=4\pi\hbar^2(a_2-a_0)/3M$. $F_{\gamma=x,y,z}$ are spin-1 matrices.
Assuming the external magnetic field $\vec B$ to be along the
quantization axis
($\hat z$), the Zeeman shift on an atom in state $|i\rangle$
becomes (the Breit-Rabi formula \cite{para1})
\begin{eqnarray}
E_\pm&=&-{E_{\rm HFS}\over 8}\mp g_I\mu_IB
    -{1\over 2}E_{\rm HFS}\sqrt{1\pm \alpha+\alpha^2},\nonumber \\
E_0&=&-{E_{\rm HFS}\over 8}
    -{1\over 2}E_{\rm HFS}\sqrt{1+\alpha^2},  \nonumber
\label{br}
\end{eqnarray}
where $E_{\rm HFS}$ is hyperfine splitting, and $g_I$ is the Lande
$g$-factor
for an atom with nuclear spin ${\vec I}$. $\mu_I$ is the
nuclear magneton and $\alpha=(g_I\mu_IB+g_J\mu_BB)/E_{\rm HFS}$ with $g_J$
representing Lande $g$-factor for a valence electron with a total angular
momentum
${\vec J}$. $\mu_B$ is the Bohr magneton.

The field operators $\psi_i$ evolve according to the
Heisenberg operator equation of motion.
At near-zero temperature and when the total number of condensed atoms
($N$) is
large, the condensate is essentially described by the mean field
$\phi_i=\langle\psi_i\rangle$. Neglecting quantum fluctuations, they form
a set of coupled Gross-Pitaevskii (GP) equations, from which we can simulate
the mean field off-equilibrium dynamics more accurately
at various external magnetic fields without using the SMA.

Our simplified model is based on the well-known fact that for both
$^{87}$Rb (ferromagnetic) and $^{23}$Na (antiferromagnetic) atoms, the spin
dependent interaction $\propto|c_2|$ is much weaker than the density
dependent interaction $\propto|c_0|$. This leads to the validity of the SMA,
where we adopt the mode function $\phi(\vec r)$ as
determined from the spin-independent part of the Hamiltonian
${\cal H}_s =-{(\hbar^2/ 2M)}\nabla^2+V + c_0n$ \cite{Law98, Pu99, Yi02}.
Thus we define
\begin{eqnarray}
\phi_i(\vec r,t) &=& \sqrt N \xi_i(t)\phi(\vec r)\exp(-i\mu t/\hbar),
\end{eqnarray}
where ${\cal H}_s\phi(\vec r)=\mu\phi(\vec r)$ and $\int d\vec r
|\phi(\vec r)|^2 = 1$. We arrive at
the coupled spinor equations
\begin{eqnarray}
i\hbar \dot{\xi_\pm} &=&
    E_\pm\xi_\pm + c[(\rho_\pm + \rho_0- \rho_\mp)\xi_\pm
     + \xi_0^2\xi_\mp^*],    \nonumber\\
i\hbar \dot{\xi_0} &=&
    E_0\xi_0 + c[(\rho_+ +\rho_-)\xi_0
     + 2\xi_+\xi_-\xi_0^*], \label{eqsma}
\end{eqnarray}
with
$c = c_2N\int d\vec r |\phi(\vec r)|^4$, $\rho_i=|\xi_i|^2$.
It is easy to verify that the total atom number and atomic
magnetization are conserved, i.e.
$\sum_i \rho_i \equiv 1$,
$\rho_+-\rho_- \equiv m$, and $m=(N_+-N_-)/N$ is a constant of motion.

We use $\eta=(E_--E_+)/2$ and
$\delta=(E_-+E_+-2E_0)/2$ to parameterize
the linear and quadratic Zeeman effect.
We further transform
\begin{eqnarray}
\xi_+ &\rightarrow & \xi_+ \exp[-i(E_0-\eta) t/\hbar], \nonumber\\
\xi_0 &\rightarrow & \xi_0 \exp[-iE_0t/\hbar], \nonumber\\
\xi_- &\rightarrow & \xi_- \exp[-i(E_0+\eta) t/\hbar] \nonumber
\end{eqnarray}
to eliminate the $E_0$ and $\eta$ dependence, and take
$\xi_j=\sqrt{\rho_j} e^{-i\theta_j}$. After some simplification,
we obtain the following dynamic equations for spin mixing
inside a spin-1 condensate
\begin{eqnarray}
\dot \rho_0 &=& {2c\over \hbar}\rho_0 \sqrt{(1-\rho_0)^2-m^2} \sin \theta,
\label{eqn0}\\
\dot \theta &=& -{2\delta\over \hbar} + {2c\over \hbar}(1-2\rho_0)
\nonumber \\
    &+& \left({2c\over \hbar}\right)
    {(1-\rho_0)(1-2\rho_0)-m^2 \over \sqrt{(1-\rho_0)^2-m^2}} \cos\theta,
\label{eqtheta}
\end{eqnarray}
where $\theta = \theta_++\theta_--2\theta_0$ is the relative phase.
These two coupled equations give rise to a classical dynamics of
a nonrigid pendulum, whose energy functional (or Hamiltonian)
can also be derived within the SMA as in \cite{Yi02}
\begin{eqnarray}
{\cal E} &=& c\rho_0\left[(1-\rho_0)+\sqrt{(1-\rho_0)^2-m^2}\cos\theta\right]
\nonumber \\
&&+\delta (1-\rho_0). \label{eng}
\end{eqnarray}
It is easy to check that $\dot \rho_0 = -(2/\hbar)\partial {\cal E}/\partial
\theta$ and $\dot \theta = (2/\hbar) \partial {\cal E}/\partial \rho_0$.

\begin{figure}
\includegraphics[width=3.25in]{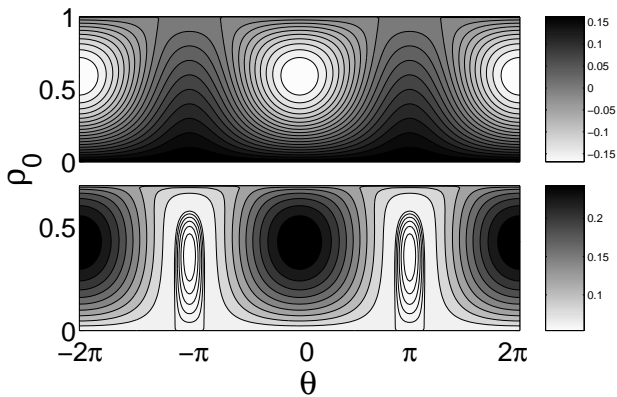}
\caption{Equal-energy contours for a condensate of $^{87}$Rb atoms (upper
panel) with $B=0.05$ Gauss, $|c|=(2\pi)0.5$ Hz, and $m=0$; of $^{23}$Na atoms
(lower panel) with $B=0.015$ Gauss, $|c|=(2\pi)0.5$ Hz, and $m=0.3$.}
\label{fig2}
\end{figure}

\begin{figure}
\includegraphics[width=3.25in]{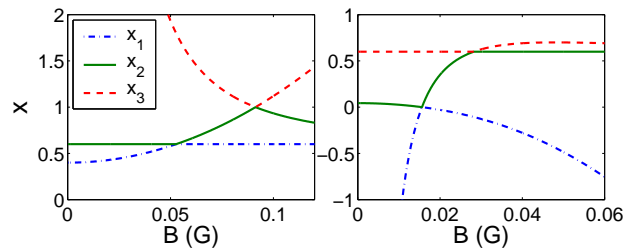}
\caption{(Color online) The dependence of cubic roots $x_j$ on the external
magnetic field for $^{87}$Rb atoms (left) and $^{23}$Na atoms (right). Other
parameters are $|c|=(2\pi) 0.5$ Hz, $\rho_0(0)=0.6$, $\theta(0)=0$, and $m=0$
for $^{87}$Rb; $\theta(0)=\pi$ and $m=0.3$ for $^{23}$Na.} \label{fig3}
\end{figure}

The contour plot of ${\cal E}$ in Fig. \ref{fig2} displays several types of
oscillation as in a pendulum. The dynamics of spin mixing described by Eqs.
(\ref{eqn0}, \ref{eqtheta}) in a magnetic field is conservative, as also
recognized and studied numerically in Ref. \cite{Romano04}. The corresponding
phase space trajectory is therefore confined to stay on the equal energy
contour. Quite generally, $\rho_0$ oscillates in a magnetic field. Rewriting
equation (\ref{eqn0}) as
\begin{eqnarray}
(\dot \rho_0)^2 &=& {4\over \hbar^2} \{ [{\cal E}-\delta(1-\rho_0)]
    [(2c\rho_0+\delta)(1-\rho_0)-{\cal E}]\nonumber \\
    &&\quad\quad-(c\rho_0m)^2\},
\end{eqnarray}
we can compute the oscillation period according to
\begin{subequations}
\begin{eqnarray}
T &=& \oint {1\over \dot \rho_0} d \rho_0
   =  {\sqrt 2\hbar \over \sqrt{-\delta c}}
        {K\left(\sqrt{{x_2-x_1 \over x_3-x_1}}\right)
        \over \sqrt{x_3-x_1}}, \hskip 2pt {\rm for\ } c<0,\ \ \
\label{RbTsma}
\end{eqnarray}
and
\begin{eqnarray}
T &=& {\sqrt 2\hbar \over \sqrt{\delta c}}
        {K\left(\sqrt{{x_3-x_2 \over x_3-x_1}}\right)
        \over \sqrt{x_3-x_1}}, \hskip 6pt {\rm for\ } c>0.
\label{NaTsma}
\end{eqnarray}
\end{subequations}
$K(k)$ is the elliptic integral of the first kind, and $x_{j=1,2,3}$ are the
roots of $\dot \rho_0=0$ (order as $x_1\le x_2\le x_3$) (Fig. \ref{fig3}). The
period for a rigid pendulum, described by $\ddot u + \sin u = 0$, is $T=4\sqrt
2K[\sqrt{2/(E+1)}\;]/\sqrt{E+1}$ at an energy $E>1$ and $T=4\sqrt
2F[\arcsin(\sqrt{(E+1)/2}\,),\sqrt{2/(E+1)}\,]/\sqrt{E+1}$ when $-1\le E \le
1$. Here $E$ is the energy of the rigid pendulum.

\begin{figure}
\includegraphics[width=3.25in]{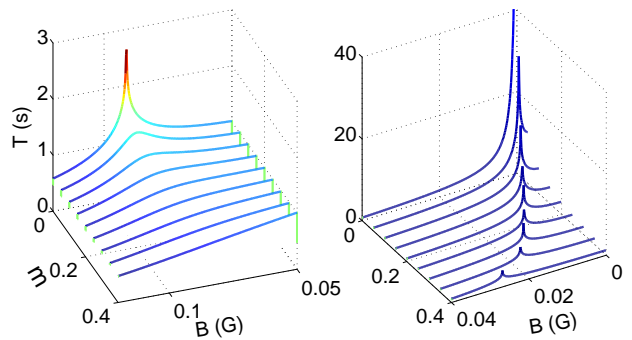}
\caption{(Color online) The magnetic field dependence of the oscillation period
for $^{87}$Rb atoms (left) and $^{23}$Na atoms (right). Other parameters are
the same as in Fig. \ref{fig3}.} \label{fig4}
\end{figure}

The time evolution of $\rho_0$ can be expressed
in terms of the
Jacobian elliptic function cn(.,.),
\begin{subequations}
\begin{eqnarray}
\rho_0(t) &=& x_2-(x_2-x_1){\rm cn}^2\left[\gamma_0
    + t\sqrt{-2\delta c(x_3-x_1)}, k\right], \nonumber\\
    && {\rm for}\ \ c<0,
\end{eqnarray}
and
\begin{eqnarray}
\rho_0(t) &=& x_3-(x_3-x_2){\rm cn}^2\left[\gamma_0
    + t\sqrt{2\delta c(x_3-x_1)}, k\right], \nonumber\\
    && {\rm for}\ \ c>0,
\end{eqnarray}
\end{subequations}
$\gamma_0$ depends on the initial state, ${\rm cn}^2 (\gamma_0,k) =
[x_2-\rho_0(0)]/(x_2-x_1)$ if $c<0$ and ${\rm cn}^2 (\gamma_0,k) =
[x_3-\rho_0(0)]/(x_3-x_2)$ if $c>0$. For $^{87}$Rb atoms ($c<0$), $\gamma_0=0$
if $\rho_0(0)=x_1$ and $\gamma_0=K(k)$ if $\rho_0(0)=x_2$. The solutions of
$\rho_0$ are oscillatory between $x_1$ and $x_2$ if $c<0$ (between $x_2$ and
$x_3$ if $c>0$), except when $x_2=x_3$ ($x_2=x_1$ if $c>0$), where the solution
becomes homoclinic, i.e., $\lim_{t\rightarrow \infty} \rho_0 = 1$ and the
corresponding period is infinity for $m=0$ (Fig. \ref{fig4}).

We further observe from Fig. \ref{fig4} that when the total magnetization is
varied the peak of the oscillation period essentially stays at the same
magnetic field for ferromagnetic interactions. The solution becomes periodic
when $m\neq 0$ since $\rho_0$ can at most reach $1-m$. It turns out that the
critical solution of an infinitely long oscillation period occurs when
$\rho_0(t\rightarrow \infty)= 1$, or equivalently ${\cal E}=0$, which gives
$\delta(B_c)=|c|\rho_0(1+\cos\theta)$ with $\rho_0$ and $\theta$ the initial
conditions. At $B=0$ we reproduce the same result as in Ref. \cite{Pu99}. The
rapid decreasing of the period when $B>B_c$ is consistent with the recent
numerical simulations by Schmaljohann {\it et al}. \cite{Schmaljohann04a}. For
antiferromagnetic interactions, however, the peak of the oscillation period
shows a strong dependence on the magnetization, and asymptotically we find
$\rho_0(t\rightarrow \infty)=0$, i.e., ${\cal E}=\delta(B_c)$ which is
equivalent to $\delta(B_c) = c[(1-\rho_0)+\sqrt{(1-\rho_0)^2-m^2}\cos\theta]$.

Substituting the solution $\rho_0(t)$ into Eq. (\ref{eng}), we can solve for
$\theta(t)$. Furthermore we can find the time dependence of $\theta_\pm$ and
$\theta_0$ through the following
\begin{eqnarray}
\dot \theta_\pm &=& -{1\over \hbar} [\delta + c\rho_0 +
    c\rho_0 \sqrt{1-\rho_0\mp m\over 1-\rho_0\pm m}\cos \theta],
\nonumber \\
\dot \theta_0 &=& -{c\over \hbar} [(1-\rho_0) +
\sqrt{(1-\rho_0)^2-m^2}\cos \theta]. \nonumber
\end{eqnarray}

Finally, we consider the evolution of the averaged total spin. As was recently
demonstrated by Higbie {\it et al.}, the averaged spin of a condensate or its
magnetization can be directly probed with non-destructive phase contrast
imaging \cite{Higbie05}. Alternatively, the magnetization dynamics can be
inferred from component populations of a spinor condensate, which are directly
measurable using Stern-Gerlach effect in an inhomogeneous magnetic field. We
first illustrate the quadratic Zeeman effect on the spin dynamics of a
noninteracting condensate. For a state ${\vec \xi}=(\xi_+, \xi_0, \xi_-)^T$,
the total spin average is $\langle \vec F\rangle = \langle {\vec \xi} | F_x\hat
x + F_y\hat y + F_z\hat z |{\vec \xi} \rangle$ with
\begin{eqnarray}
\langle F_x \rangle &=& \sqrt 2 {\rm Re}
{\left[|\xi_0|\left(|\xi_+|e^{i(\theta_0-\theta_+)} +
|\xi_-| e^{i(\theta_0-\theta_-)}\right)\right]}, \nonumber \\
\langle F_y \rangle &=& \sqrt 2 {\rm Im} {\left[|\xi_0|\left(|\xi_+|
e^{i(\theta_0-\theta_+)} -
|\xi_-| e^{i(\theta_0-\theta_-)}\right)\right]}, \nonumber \\
\langle F_z \rangle &=& |\xi_+|^2-|\xi_-|^2 = m. \nonumber
\end{eqnarray}
As an interesting case, we take the initial state as ${\vec \xi}(0) =
\left[\sqrt{(1-\rho_0)/2}, \sqrt {\rho_0}, \sqrt{(1-\rho_0)/2}\:\right]^T$.
$\rho_0$ is a constant. We find at time $t$ that
\begin{eqnarray}
\langle F_x \rangle+i\langle F_y \rangle &=& 2\sqrt{\rho_0(1-\rho_0)}
\cos(\delta t/\hbar)\, e^{-i\eta t/\hbar},\nonumber \\
\langle F_z \rangle &=& m = 0.
\end{eqnarray}
It spirals toward and away from the origin in the
$\langle F_x \rangle$-$\langle F_y\rangle$ plane.
The linear Zeeman effect causes spin precessing around
the magnetic field ($\hat z$ axis), while the quadratic Zeeman effect
makes spin average oscillate.

\begin{figure}
\includegraphics[width=3.25in]{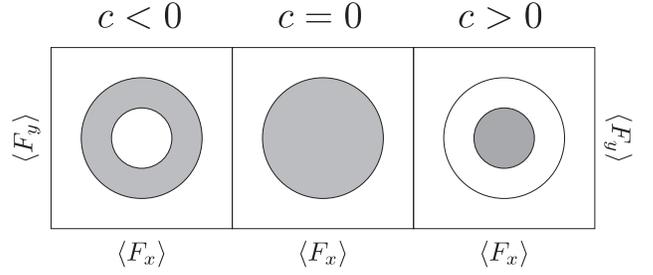}
\caption{Two dimensional projection of the averaged spin evolution (shaded
region) for a condensate with zero magnetization of noninteracting atoms
(middle), in comparison with atoms of ferromagnetic (left) and
antiferromagnetic interactions (right).} \label{fig5}
\end{figure}

The spin evolution becomes quite different when atom interaction is present.
For the same initial conditions (of the above), the total averaged spin
at time $t$ becomes
\begin{eqnarray}
\langle F_x\rangle+i\langle F_y\rangle &=&
2\sqrt{\rho_0(1-\rho_0)}\cos(\theta/2)\, e^{-i\eta t/\hbar},\nonumber \\
\langle F_z \rangle &=& 0,
\end{eqnarray}
which can be conveniently confirmed from the phase space contour plot of Fig.
\ref{fig2}, where $\theta$ is confined to oscillate around zero for
ferromagnetic interactions and around $\pi$ for antiferromagnetic interactions
if $B<B_c$. Note that $\rho_0$ and $\theta$ are time-dependent for interacting
condensates. Figure \ref{fig5} exemplifies this oscillation in terms of the
allowed regions (shaded) of $\langle F_x\rangle$ and $\langle F_y\rangle$ for
interacting condensates in contrast to non-interacting ones. For ferromagnetic
interactions, the allowed region is defined by two radii. One of them,
$r_I=\sqrt{2\rho_0(0)[(1-\rho_0(0))+ \sqrt{(1-\rho_0(0))^2-m^2}\;]}$, depends
on the initial condition, while the other ($r_B$) is solely determined by the
quadratic Zeeman effect. We find $r_B>r_I$ if $B<B_0$, $0<r_B<r_I$ if
$B_0<B<B_c$, and $r_B=0$ if $B\ge B_c$. There exists a forbidden region at the
center for a ferromagnetically interacting condensate if $B<B_c$. This region
shrinks to zero when $B\ge B_c$. Exactly at $B=B_c$, an interesting
attractor-like feature arises and the average spin gradually spirals towards
the origin (at the center) and becomes trapped eventually after an infinitely
long time. For antiferromagnetic interactions, the allowed region generally
becomes smaller than that for a noninteracting condensate as shown in Fig.
\ref{fig5} for $m=0$ (or $B_c=0$). The radius of the shaded (allowed) region
depends on the quadratic Zeeman effect, while the forbidden region approaches
zero as $B\to \infty$.

For the general case of $m\neq 0$, the allowed region is in between the two
radii $\sqrt{2\rho_0(0)[(1-\rho_0(0))\pm \sqrt{(1-\rho_0(0))^2-m^2}\;]}$ for a
noninteracting gas, For ferromagnetic interactions, the averaged spin behaves
similar to the case of $m=0$ considered above, except now the forbidden region
shrinks gradually to a minimum nonzero value of
$\sqrt{2\rho_0(0)[(1-\rho_0(0))-\sqrt{(1-\rho_0(0))^2-m^2}\;]}$ when $B\to
\infty$. In this case, there exists no $B_c$ or homoclinic orbits. For
antiferromagnetic interactions, the analogous radius $r_B$ decreases from
$r_I=\sqrt{2\rho_0(0)[(1-\rho_0(0))-\sqrt{(1-\rho_0(0))^2-m^2}\;]}$ to zero
while $B$ increases from zero to $B_c$. At $B=B_c$ the attractor-like feature
remains. When $B$ is increased from $B_c$, $r_B$ increases from zero, and
crosses $r_I$ at $B=B_0$, finally approaches the radius of the allowed region
for a
non-interacting condensate when $B\to \infty$. \\

Before concluding, we hope to make some estimates to support the use of the
mean field theory, i.e. treating the atomic field operators as c-numbers.
Intuitively, we would expect that this is a reasonable approximation as the
total numbers of atom, at 1000, although not macroscopic, is definitely
`large'. In fact, the recent double well experiment that confirmed the coherent
nonlinear Josephson oscillations of the mean field theory, is at a similar
level of numbers of atoms \cite{Albiez04}. A rigorous discussion of this point
in terms of the quantum phase diffusions in a spin-1 condensate is a rather
involved procedure, and will not be reproduced here \cite{Yi03}. Instead, we
illuminate the validity of mean field theory as follows. First, we look at the
total atom number fluctuations. Approximating the spinor condensate as a one
component scalar, and neglecting the internal spin mixing dynamics, its total
overall phase spreads after a time of $\tau_c\approx N/[\sigma(N) (c_0\langle n
\rangle)]$ \cite{Imamoglu97}, with $\sigma(N)\sim \sqrt{N}$ the standard
deviation of the atom numbers from taking c-number approximations of the atomic
field operators. In our case, this time is about $0.2$ second, short compared
to a typical Josephson type oscillation period at $\sim 1$ second. We believe,
however, this is not a critical issue as we are not studying phase sensitive
phenomena involving the overall phase as in an interference experiment.
Instead, we are interested here in the relative phase dynamics between
different condensate components, whose oscillation time scale is given by the
much smaller value of the spin-dependent interaction coefficient $c_2$; thus we
should compare the coherent classical oscillation period of $\sim 1$ second
with the much longer time $\tau_c^\prime\approx N/[\sigma(N) (c_2\langle
n\rangle)]$ \cite{Law98}, $\sim 50$ seconds (for $^{87}$Rb). This then leads to
a favorable condition for adopting the mean field theory in our study.
Alternatively, we can reach the same conclusion from a direct investigation of
the oscillation period $T$ Eq. (\ref{RbTsma}), which contains a simple $N$
dependence $\propto 1/\sqrt N$. We find that $|T(N\pm\sqrt{N})-T(N)|/T(N) =
1/(2\sqrt N)$, is only about $2\%$, indicating the overall validity of the mean
field theory.

In conclusion we have studied the off-equilibrium interaction
driven collective oscillations inside an atomic condensate in an
external uniform magnetic field. The dynamics of spin mixing
is found to be well described by a nonrigid pendulum
due to the conservation of atom numbers and atomic
magnetization. In particular, we find that there exists
an interesting class of critical trajectories
whose oscillation periods approach infinity.
Our study illuminates the use of quadratic Zeeman
shift to probe pendulum-like oscillations
in a spin-1 condensate and provides the complete spin mixing dynamics
analytically.
It provides the much needed theoretical guidance for the eventual
experimental detection of coherent macroscopic oscillations
in a spinor condensate.


{\it Note added in proof:} We have recently observed many of the coherent
oscillatory behavior discussed in this work and have submitted a publication
describing these experiments.

W. Zhang thanks Dr. Su Yi for several discussions of this project. This work is
supported by NSF and NASA.

\end{document}